\documentclass[preprint,showpacs,preprintnumbers,amsmath,amssymb,prb]{revtex4}
\usepackage{graphicx}
\usepackage{dcolumn}
\usepackage{bm}
\usepackage{epsfig}
\usepackage{subfigure}
\begin{document}

\preprint{PREPRINT}

\title{Thermodynamic, Dynamic, Structural and Excess Entropy Anomalies for 
core-softened potentials}

\author{Ney M. Barraz Jr.}
\affiliation{Instituto de F\'{\i}sica, Universidade Federal do Rio Grande 
do Sul, 91501-970, Porto Alegre, RS, Brazil}

\author{Evy Salcedo}
\affiliation{Instituto de F\'{\i}sica, Universidade Federal de 
Santa Catarina, 88010-970, Florian\'opolis, SC, Brazil}

\author{Marcia C. Barbosa}
\affiliation{Instituto de F\'{i}sica, Universidade Federal do Rio 
Grande do Sul, 91501-970, Porto Alegre, Rio Grande do Sul}

\date{\today}
\begin{abstract}
Using molecular dynamic simulations we study three families of
continuous core-softened potentials consisting of two length scales: a
shoulder scale and an attractive scale. All the families have the same
slope between the two length scales but exhibit different potential
energy gap between them. For each family three shoulder depths are
analyzed.  We show that all these systems exhibit a liquid-liquid
phase transition between a high density liquid phase and a low density
liquid phase ending at a critical point. The critical temperature is
the same for all cases suggesting that the critical temperature is
only dependent on the slope between the two scales.  The critical
pressure decreases with the decrease of the potential energy gap
between the two scales suggesting that the pressure is responsible for
forming the high density liquid.  We also show, using the radial
distribution function and the excess entropy analysis, that the
density, the diffusion and the structural anomalies are present if
particles move from the attractive scale to the shoulder scale with
the increase of the temperature indicating that the anomalous behavior
depends only in what happens up to the second coordination shell.
\end{abstract}
\pacs{64.70.Pf, 82.70.Dd, 83.10.Rs, 61.20.Ja}
\maketitle

\section{\label{sec1}Introduction}

The phase behavior of single component systems as particles
interacting via the so-called core-softened $(CS)$ potentials is
receiving a lot of attention recently.  These potentials exhibit a
repulsive core with a softening region with a shoulder or a
ramp~\cite{He70,St72,Si76,Ja99a,Ja99b,Wi02,Ca03,Vi11}. These 
models originates from the desire of
constructing a simple two-body isotropic potential capable of
describing the complicated features of systems interacting via
anisotropic potentials. This procedure 
generates systems that are
analytically and computationally tractable and that one hopes are
capable to retain the qualitative features of the real complex
systems~\cite{Ke75,An76,Ts91,An00}.

The physical motivation behind these studies is the recently
acknowledged possibility that some single component systems
interacting through a core-softened potential display density and
diffusion anomalies. This opened the discussion about the relation
between the existence of thermodynamic anomalies in liquids and the
form of the effective potential~\cite{Ya08,Ku08,Ol08b,Ol09,Eg08a,Ba09}.

These anomalies appear in two different ways. First, it is the density
anomaly.  Most liquids contract upon cooling. This is not the case of
water and other fluid systems. For water the specific volume at
ambient pressure starts to increase when cooled below $T\approx4 ^oC$.
The anomalous behavior of water was first suggested 
300 years ago~\cite{Wa64} and was  confirmed by a number
of experiments~\cite{Ke75,An76}.
Besides, between $0.1$  MPa and $190$ MPa water also 
exhibits an anomalous increase of 
compressibility~\cite{Sp76,Ka79} and, at atmospheric pressure, an 
increase of isobaric
heat capacity upon cooling~\cite{An82,To99}.

Experiments for Te,
\cite{Th76} Ga, Bi,~\cite{Handbook} S,~\cite{Sa67,Ke83} and
Ge$_{15}$Te$_{85}$,~\cite{Ts91} and simulations for
silica,~\cite{An00,Ru06b,Sh02,Po97} silicon~\cite{Sa03} and
BeF$_2$,~\cite{An00} show the same density anomaly.

But density anomaly is not the only unusual behavior that these
materials have.  For a normal liquid the diffusion constant, $D$,
decreases under compression. This is not the case of water. $D$
increases on compression at low temperature, $T$, up to a maximum
$D_{\rm max} (T)$ at $p = p_{D\mathrm{max}}(T)$.~\cite{An76,Pr87}.
Numerical simulations for SPC/E water~\cite{spce} recover the
experimental results and show that the anomalous behavior of $D$
extends to the metastable liquid phase of water at negative pressure,
a region that is difficult to access for
experiments~\cite{Ne01,Er01,Mu05,Mi06a}. In this region the
diffusivity $D$ decreases for decreasing $p$ until it reaches a
minimum value $D_{\rm min} (T)$ at some pressure
$p_{D\mathrm{min}}(T)$, and the normal behavior, with $D$ increasing
for decreasing $p$, is reestablished only for $p <
p_{D\mathrm{min}}(T)$~\cite{Ne01,Ne02a,Ne02b,Er01,Mu05,Mi06a}. Besides
water, silica~\cite{Sh02,Ru06b} and silicon~\cite{Mo05} also exhibit a
diffusion anomalous region.

Acknowledging that CS potentials may engender density and diffusion
anomalous behavior, a number of CS potentials were proposed to model
the anisotropic systems described above.  They possess a repulsive
core that exhibits a region of softening where the slope changes
dramatically. This region can be a shoulder or a
ramp~\cite{He70,Sc00,Bu02,Bu03,Sk04,Fr02,Ba04,Ol05,He05a,He05b,
  Ja98,Wi02,Ma04,Ku04,Xu05,Ol06a,Ol06b,Ol07,Ol08a,Ol08b,Ol09,Gr09,Lo07}.
These models exhibit density and diffusion anomalies, but depending on
the specific shape of the potential, the anomalies might be hidden in
the metastable phase  \cite{Ol09}. Also there are a
number of core-softened potentials in which the anomalies are not
present \cite{Fr01,Si10}. The relation between the specific shape of
the effective core-softened potential and the microscopic mechanism
necessary for the presence of the anomalies is still under debate 
\cite{Ya06,Ol08a,Ol09,Vi10}.

Recently it was suggested that the link between the presence of the
density and diffusion anomaly and the microscopic details of the
system can be analyzed in the framework of the excess-entropy-based
formalism \cite{Ba89} applied to similar systems by Errington \emph{et
  al.}  \cite{Er06} and Chakraborty and Chakravarty \cite{Cha06}.
Within this approach the presence of the density and the diffusion
anomalies are related to the density dependence of the excess entropy,
$s_{\mathrm{ex}}$.

The computation of the excess entropy, however, requires integrating
the radial distribution function in the whole space.
The anomalous behavior, however, seems to depend only in the two
length scales present in the system and, therefore, should not depend
on the particle distributions far away.  Here we propose that 
two length scales potentials will have density and
diffusion anomalies if the two  scales would 
be accessible. In principle the accessibility only
depends in the distribution of particles in these 
two distances. Therefore, the
knowledge of the complete excess entropy is not necessary for knowing
if the system has or not anomalies.  The behavior of the partial
excess entropy, computed only up to the second coordination shell
should give enough information to determine if a system has
anomalies or not.

In this paper we test this assumption by
computing the pressure-temperature phase 
diagram and the 
excess entropy for  three families of
core-softened potentials that have two length scale: a shoulder scale
and an attractive scale \cite{He93,Ba09}.  In all the 
three families the slope between
the two scales is the same~\cite{Ya08}. The shoulder 
scale is made more favorable by decreasing the 
energy gap between the two length scales (see potentials
A, B and C in Fig.~\ref{fig:potential2}). In addition the
shoulder scale becomes more favorable by 
 making the depth of 
the shoulder scale deeper (see potentials
A1, A2 and A3 in Fig.~\ref{fig:potential2}). The slope between
the two length scales is kept
fixed in order to have the liquid-liquid critical point and the density
anomalous region in the same region of the pressure temperature phase
diagram~\cite{Ya08}.  

The remaining of this paper goes as follows. In Sec.~\ref{sec:model}
the model is introduced.  The simulations details are given
Sec.~\ref{sec:simulation}.  In Sec.~\ref{sec:results} the results are
discussed. Finally, Sec.~\ref{sec:conclusions} presents the
conclusion.

\section{\label{sec:model} The Model}

We consider a system of $N$ particles, with diameter $\sigma$, 
where the pair interaction is described by a family of continuous
potentials given by
\begin{equation}\label{eq:potential}
 \centering
  U(r) = \epsilon \left[ \left( \frac{\sigma}{r} \right)^{a} - 
\left( \frac{\sigma}{r} \right)^{b} \right] + \sum_{j=1}^{4}h_{j}
  \exp \left[ -\left( \frac{r-c_{j}}{w_{j}} \right)^{2} \right] \;\;.
\end{equation}

\begin{figure}[ht]
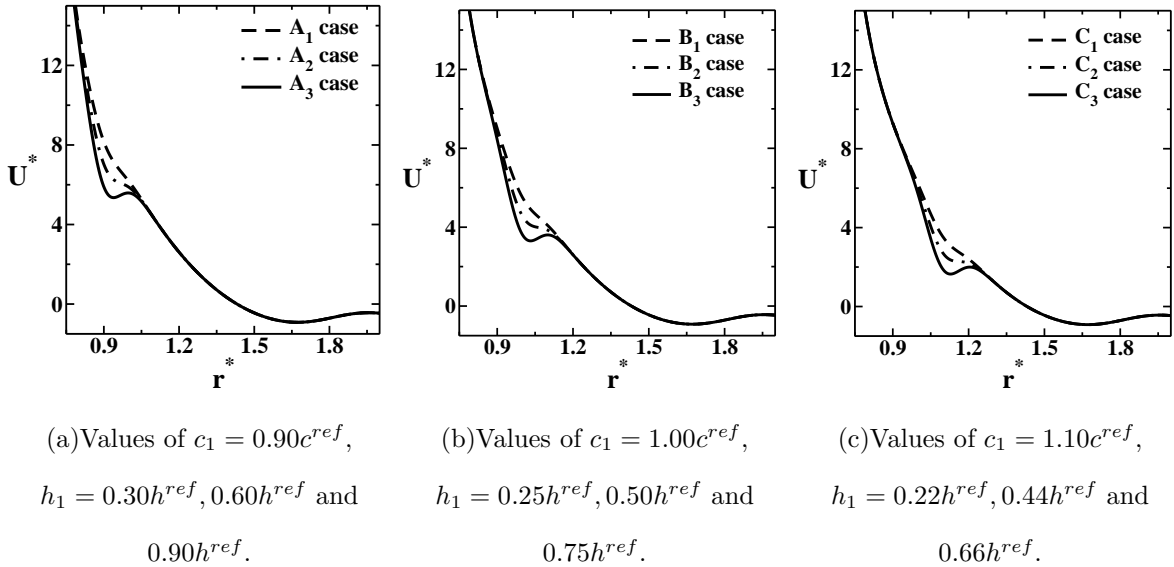

  \begin{centering}
    \subfigure[ Values of $c_1 = 0.90 c^{ref}$, $h_1 = 0.30h^{ref}, 0.60h^{ref}$ and $0.90h^{ref}$.]{\includegraphics[clip=true,width=5cm]{potential_A.eps} }
    \subfigure[ Values of $c_1 = 1.00 c^{ref}$, $h_1 = 0.25h^{ref}, 0.50h^{ref}$ and $0.75h^{ref}$.]{\includegraphics[clip=true,width=5cm]{potential_B.eps} }
    \subfigure[ Values of $c_1 = 1.10 c^{ref}$, $h_1 = 0.22h^{ref}, 0.44h^{ref}$ and $0.66h^{ref}$.]{\includegraphics[clip=true,width=5cm]{potential_C.eps} }
    \par
  \end{centering}
  \caption{Interaction potential obtained by changing parameters $h_1$ in the
    Eq.~(\ref{eq:potential}).The potential and the 
distances are in dimensionless
units $U^*=U/\gamma$ and $r^*=r/r_0$.Here we use $\epsilon/\gamma=0.02$ and
$\sigma/r_0=1.47$.}
  \label{fig:potential2}
\end{figure}

\begin{figure}[ht]
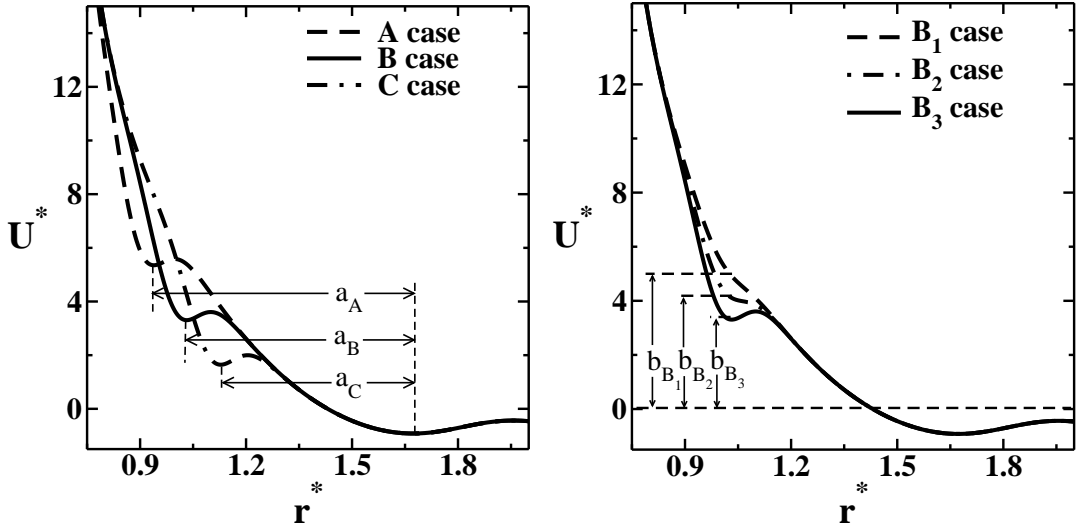

  \begin{centering}
    \subfigure[\ Distance between the scales.]{{\includegraphics[clip=true,width=7cm]
{potential_a.eps}} \label{fig:potential_a}}
    \subfigure[\ Different depths of the first scale.]{{\includegraphics[clip=true,width=7cm]
{potential_c.eps}} \label{fig:potential_b}}
    \par
  \end{centering}
  \caption{Scheme to distinguish the distance between the scales and depths of the first scale.}
\label{fig:potentialABC}
\end{figure}

The first term is a Lennard-Jones potential and the second term is
composed by four Gaussians, each Gaussian is centered in $c_j$. This
potential can represent a whole family of intermolecular interactions,
depending of the choice of the parameters $a,b,\sigma,\{h_j, c_j,
w_j\}$, with $j=1, \dots, 4$. The values of these parameters are in
Table~\ref{table:ref}. The parameters are chosen in order to obtain a
two length scale potential~\cite{He93,St93,De91} that is related to the
interaction between two tetramers~\cite{He93,St93}.

The simulations are made in dimensionless units, therefore all the
physical quantities are given in terms of the energy scale $\gamma$
and the distance scale $r_0$ where $\gamma$ is the  energy scale and
$r_0$ is the length scale chosen so the minimum of the 
potential in the $B_3$ case 
is about $r^*=1$.  The distance
scale is chosen so the minimum of the shoulder
scale in the $B_3$ case  Here we use $\epsilon/\gamma=0.02$ and
$\sigma/r_0=1.47$.

Modifying the parameters $c_1$ and $h_1$ in the
Eq.~(\ref{eq:potential}), according to Table~\ref{table:shoulder},
allow us to change the depth of the shoulder well, as illustrated in
Fig.~\ref{fig:potential2}. Here we use nine 
different values for $h_1$ and they are expressed as a multiple of a reference 
value $h_1^{ref}$.  We also use three different values
of $c_1$  and they are expressed as a multiple of a reference 
value $c_1^{ref}$. For all the nine cases 
the values of $a,b,\{w_j\}$ with $j=1, \dots, 4$, ${c_i}$
with $i=2, \dots, 3$, $h^{ref}$ and $c^{href}$. Table~\ref{table:ref} 
gives the parameter values in $\AA$ and $kcal/mol$ 
consistent with modeling ST4 water~\cite{He93}.

Modifying the distance between the two minima of the two scales,
shoulder scale $(SC)$ and attractive scale $(AS)$, leads to the three
families $A$, $B$ and $C$ as shown in Fig.~\ref{fig:potential2}.  The
changes in the distance between the two length scales were done in
such way to preserve the slope between the two scales and, therefore,
to have in all the cases the region of density anomaly in the same
region of the pressure-temperature phase diagram as proposed by Yan
et. al~\cite{Ya07}.

The family $A$ has the largest distance and the largest potential
energy gap between the two length scales ($a_A=0.72$ in
Fig.~\ref{fig:potentialABC}), the family $B$ has the intermediate
distance and intermediate potential energy gap between the two length
scales ($a_B=0.62$ in Fig.~\ref{fig:potentialABC}), and the family $C$
has the shortest distance and the smallest potential energy gap
between the two length scales, ($a_C=0.52$ in
Fig.~\ref{fig:potentialABC}). 

For each family, we analyze three different depths of the shoulder
scale represented by $1$, $2$ and $3$ (see
Fig.~\ref{fig:potentialABC}). The potentials $1$ have the most shallow
shoulder scale , the potentials $2$ have intermediate shoulder depth
and the potentials $3$ have the deepest shoulder scale.
Table~\ref{table:b} gives the values for the depths for each one of
the families.

In summary, we analyze nine different potentials: $A_1, A_2, A_3, B_1,
B_2, B_3, C_1, C_2$ and $C_3$ as illustrated in
Fig.~\ref{fig:potential2}.  The values of the different $h_1$ for each
case are listed in Table~\ref{table:shoulder}.

Barraz et~al.~\cite{Ba09} investigated the family $B$. It was shown
that this potential exhibits thermodynamic, dynamic and structural
anomalies if the shoulder scale is not too deep. Their result suggests
that in order to have anomalies it is necessary but no sufficient to
have two length scales competing.  Both length scales must be
accessible.  When the shoulder scale becomes too deep the particles
are trapped in this length scale and no anomaly is present. By making
the shoulder deeper we are decreasing the difference in energy between
the scales and therefore destroying the competition.

Here we explore the accessibility or the absence of accessibility by
changing both the energy gap between the two length scales and the
distance between them.

\section{\label{sec:simulation} The Simulation Details}

The properties of the system were obtained by $NVT$ molecular dynamics
using Nose-Hoover heat-bath with coupling parameter $Q = 2$. The
system is characterized by 500 particles in a cubic box with periodic
boundary conditions, interacting with the intermolecular potential
described above. All physical quantities are expressed in reduced
units~\cite{Al87} given by
\begin{eqnarray}
t^* &=& tr_0(m/\gamma)^{1/2}\nonumber  \\
T^* &=& \frac{k_{B}T}{\gamma }\nonumber  \\
p^*& =& \frac{pr_0^3}{\gamma }\nonumber \\
\rho^{*} &=& \rho r_0^3 \nonumber \\
D^* &=& D \sqrt{\frac{m}{\gamma r_0^2}} \;\; .
\end{eqnarray}


Standard periodic boundary conditions together with
predictor-corrector algorithm were used to integrate the equations of
motion with a time step $\Delta t^{*}=0.002$ and potential cut off
radius $r_{c}^{*}=3.5$. 
The initial configuration were  set on solid and on
liquid states and, in both cases, the equilibrium state was reached
after $t_{eq}^{*}=1000$ (what is in fact $500,000$ steps since $\Delta
t^{*}=0.002$). From this time on the physical quantities were stored
in intervals of $\Delta t_R^* = 1$ during $t_R^* = 1000$. The system
is uncorrelated after $t_d^*= 10$, from the velocity auto-correlation
function. With $50$ descorrelated samples were used to get the average
of the physical quantities.

The thermodynamic stability of the system
was checked by analyzing the dependence of pressure on density namely
\begin{eqnarray}
\frac{\partial p}{\partial \rho} < 0
\label{eq:stability}
\end{eqnarray}
, by the
behavior of the energy and also by visual analysis of the final
structure, searching for cavitation. 

At the phase boundary between the liquid and the amorphous phases
we found stable states points at both phases. The state with 
the lower energy was considered. In this particular region of 
the pressure-temperature phase diagram the energy was a good
approximation for the Helmoltz free energy.

 The error bars 
for temperatures and pressures
away from the critical region are smaller than the size of
the gray lines.  The error bar near the critical point are
$\Delta T = 0.0025$ and $\Delta p = 0.05$. 
Our error is controlled by making averages of uncorrelated measures.

\begin{center}
 \begin{table}
  \caption{Parameters for potentials A, B and C in reduced unit.}
  \begin{tabular}{cc|cc|cc|cc}
  \hline\hline

\ \ Parameter \ &  Value \ \ & \ \ Parameter \ & Value \ \ & \ \ Parameter     \ & Value \ \ & \ \ Parameter     \ & Value                   \ \ \tabularnewline \hline
\ \   $a$     \ & $9.056$  \ & \ \  $w_{1}$  \ & $0.085$ \ & \ \ $c_{1}^{ref}$ \ & $0.996$ \ & \ \ $h_{1}^{ref}$ \ &                $-3.79$ \ \ \tabularnewline
\ \   $b$     \ & $4.044$  \ & \ \  $w_{2}$  \ & $0.618$ \ & \ \ $c_{2}$       \ & $0.529$ \ & \ \ $h_{2}$       \ & \hspace{0.2cm} $1.209$  \ \ \tabularnewline
\ \ $\epsilon$\ & $0.020$  \ & \ \  $w_{3}$  \ & $0.826$ \ & \ \ $c_{3}$       \ & $1.598$ \ & \ \ $h_{3}$       \ &                $-1.503$ \ \ \tabularnewline
\ \ $\sigma$  \ & $1.475$  \ & \ \  $w_{4}$  \ & $0.214$ \ & \ \ $c_{4}$       \ & $1.929$ \ & \ \ $h_{4}$       \ & \hspace{0.2cm} $0.767$  \ \ \tabularnewline
\hline\hline

  \end{tabular}\label{table:ref}
 \end{table}
\end{center}

\begin{center}
 \begin{table}
  \caption{Parameters of $c_1$ and $h_1$ for potentials A, B and C.}
  \begin{tabular}{cc||cc|cc|cc}
  \hline\hline \vspace{-0.76cm}
\\ Potentials & Values      \ \ & \ \ Potentials \ &  Values     \ \ & \ \ Potentials \ &  Values     \ \ & \ \ Potentials \ &  Values     \ \ \tabularnewline \hline
\ \   $A$   \ & $ 0.90\, c_1^{ref}$ \ \ &\ \   $A_1$    \ & $ 0.30\, h_1^{ref}$ \ \ & \ \   $B_1$    \ & $ 0.25\, h_1^{ref}$ \ \ & \ \    $C_1$   \ & $ 0.22\, h_1^{ref}$ \ \
\tabularnewline
\ \   $B$   \ & $ 1.00\, c_1^{ref}$ \ \ &\ \   $A_2$    \ & $ 0.60\, h_1^{ref}$ \ \ & \ \   $B_2$    \ & $ 0.50\, h_1^{ref}$ \ \ & \ \    $C_2$   \ & $ 0.44\, h_1^{ref}$ \ \ \tabularnewline
\ \   $C$   \ & $ 1.10\, c_1^{ref}$ \ \ &\ \   $A_3$    \ & $ 0.90\, h_1^{ref}$ \ \ & \ \   $B_3$    \ & $ 0.75\, h_1^{ref}$ \ \ & \ \    $C_3$   \ & $ 0.66\, h_1^{ref}$ \ \ \tabularnewline
\hline\hline
  \end{tabular}\label{table:shoulder}
 \end{table}
\end{center}

\begin{center}
 \begin{table}
  \caption{Values for the depths for each one of the families.}
  \begin{tabular}{cc|cc|cc}
  \hline\hline

\ \ Potential \ &  Value \ \ & \ \ Potential \ & Value \ \ & \ \ Potential \ & Value    \ \ \tabularnewline \hline
\ \   $b_{A_1}$   \ & $7.10$  \ & \ \  $b_{B_1}$    \ & $4.94$ \ & \ \ $b_{C_1}$     \ & $2.95$  \ \ \tabularnewline
\ \   $b_{A_2}$   \ & $6.20$  \ & \ \  $b_{B_2}$    \ & $4.07$ \ & \ \ $b_{C_2}$     \ & $2.33$  \ \ \tabularnewline
\ \   $b_{A_3}$   \ & $5.28$  \ & \ \  $b_{B_3}$    \ & $3.32$ \ & \ \ $b_{C_3}$     \ & $1.60$   \ \ \tabularnewline

\hline\hline

  \end{tabular}\label{table:b}
 \end{table}
\end{center}

\section{\label{sec:results} Results}

\subsection*{\label{phase-diagram} Pressure-Temperature Phase Diagram}

\begin{figure}[!]
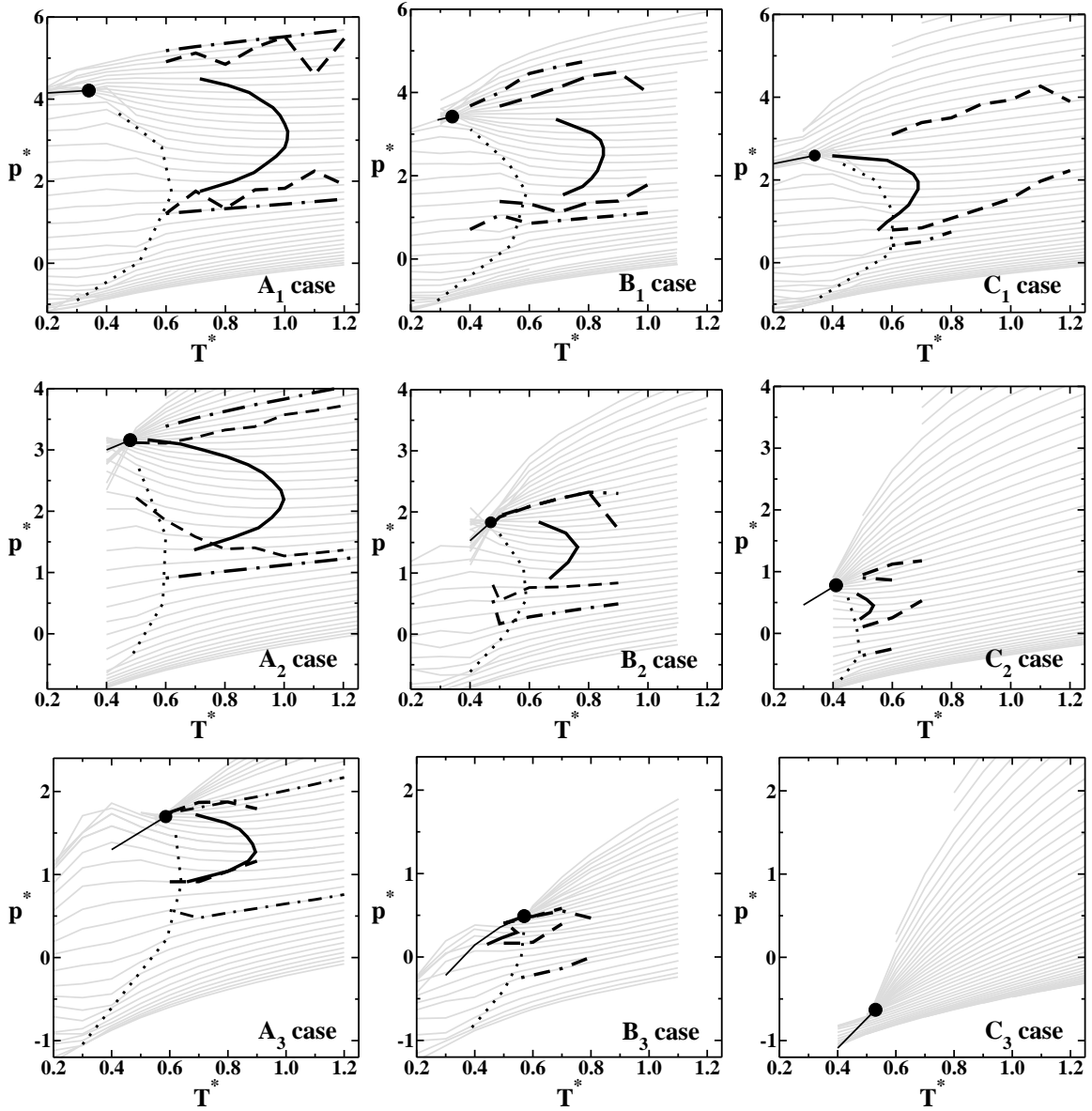

 \centering
 \begin{tabular}{ccc}
  \includegraphics[clip=true,width=5cm]{PT_A1.eps} &
  \includegraphics[clip=true,width=5cm]{PT_B1.eps} &
  \includegraphics[clip=true,width=5cm]{PT_C1.eps} \tabularnewline
  \includegraphics[clip=true,width=5cm]{PT_A2.eps} &
  \includegraphics[clip=true,width=5cm]{PT_B2.eps} &
  \includegraphics[clip=true,width=5cm]{PT_C2.eps} \tabularnewline
  \includegraphics[clip=true,width=5cm]{PT_A3.eps} &
  \includegraphics[clip=true,width=5cm]{PT_B3.eps} &
  \includegraphics[clip=true,width=5cm]{PT_C3.eps} \tabularnewline
 \end{tabular}\par
 \caption{Pressure-temperature phase diagram for the family $A$ in the
   left-hand side, for the family $B$ in the middle and for $C$ in the
   right-hand side. The thin solid lines are the isochores
   $0.30<\rho^*<0.65$. The liquid-liquid critical point is the dot,
   the temperature of maximum density is the solid thick line, the
   diffusion extrema is the dashed line and the structural extrema is
   the dashed-dotted line. The dotted line indicates the limit between
   the fluid and the amorphous regions. For the diffusion coefficient
   and radial distribution functions it is possible to determine the
   amorphous region. The full line on the left side of the second
   critical point is an approximate line of the coexistence of fluids
   of high and low density. }\label{fig:PT}
\end{figure}

First, we explore  the effects that the increase of 
the shoulder depth and  the decrease of the  distance
between the two scales have in the location in the pressure-temperature
phase diagram of the different phases. Fig.~\ref{fig:PT} illustrates
the pressure versus temperature phase diagram of the three families
$A$, $B$ and $C$ of potentials.  At high temperatures there are a
fluid phase and a gas phase (not shown). These two phases coexist at a
first order line that ends at a critical point (see
Table~\ref{table:CP_GL} for the pressure and the temperature values).

At low temperatures and high pressures there are two liquid phases
coexisting at a first order line ending at a second critical point
(see Table~\ref{table:CP_LL} for the pressure and the temperature
values).  The thermodynamic stability of the state points
was checked by analyzing the dependence of pressure on density
 using  Eq.~\ref{eq:stability}.
The critical point was identified in the graph by the region
where isochores cross. The coexistence line was obtained as the medium
line between the stability limit of each phase.

\begin{figure}[ht]
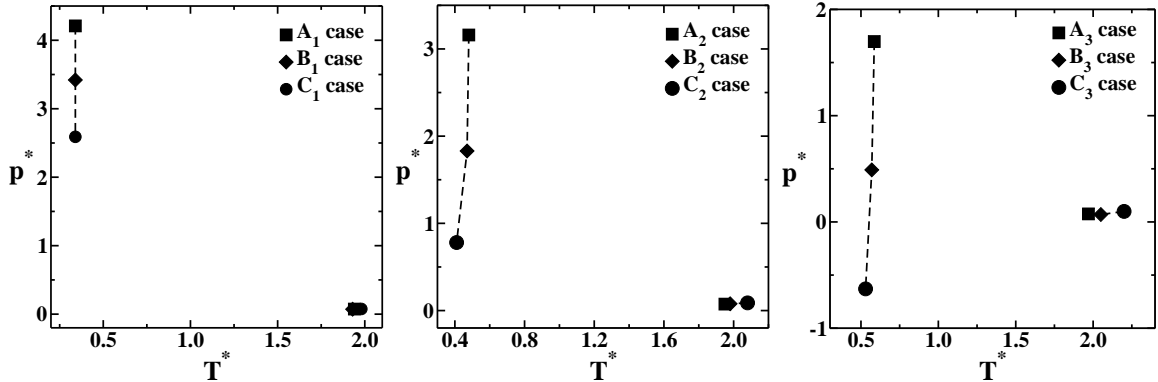

 \begin{centering}
  \begin{tabular}{ccc}
   \includegraphics[clip=true,width=5cm]{PT_CP1.eps} &
   \includegraphics[clip=true,width=5cm]{PT_CP2.eps} &
   \includegraphics[clip=true,width=5cm]{PT_CP3.eps} \tabularnewline
  \end{tabular}\par
 \end{centering}
 \caption{Location of the critical points on pressure-temperature phase diagram 
for cases $A$, $B$ and $C$. }\label{fig:PT_CP}
\end{figure}

 Comparison
between the cases $A_1$, $B_1$ and $C_1$ indicates that as the
distance between the two scales decreases, the critical pressure
decreases but the critical temperature remains the same as illustrated
in Fig.~\ref{fig:PT_CP}. This observation is confirmed in the cases
$A_2$, $B_2$ and $C_2$ and in the cases $A_3$, $B_3$ and $C_3$ (see
Table~\ref{table:CP_LL} for the critical pressures and the
temperatures). 

How can this result be understood? The two liquid phases are formed
due to the presence of the two competing scales. The low density
liquid is related to the attractive scale while the high density
liquid is related to the shoulder scale. In order to reach the high
density liquid phase the system has to overcome a large potential
energy but also have to
become very compact in the
case $A_1$. The potential energy gap 
and the distance between the two length scales are
higher in the case $A$ than in the case $C$, therefore the pressure
needed for forming a high density liquid phase is higher in the case
$A $ than it is in the case $C$.

\begin{center}
 \begin{table}
  \caption{First critical point location for potentials $A$, $B$ and $C$.}
  \centering{}
  \begin{tabular}{ccc|ccc|ccc}
\hline\hline
\ \ Potential &  $T_{c1}^*$  &  $p_{c1}^*$ \ \ & \ \ Potential & $T_{c1}^*$  &  $p_{c1}^*$ \ \ & \ \ Potential & $T_{c1}^*$  &  $p_{c1}^*$ \ \  \tabularnewline \hline
\ \  $A_1$    &   $1.94$     &  $0.074$    \ \ & \ \ $B_1$     & $1.93$      &  $0.072$    \ \ & \ \ $C_1$     &  $1.98$     &  $0.076$    \ \  \tabularnewline
\ \  $A_2$    &   $1.95$     &  $0.074$    \ \ & \ \ $B_2$     & $1.98$      &  $0.078$    \ \ & \ \ $C_2$     &  $2.08$     &  $0.088$    \ \  \tabularnewline
\ \  $A_3$    &   $1.97$     &  $0.076$    \ \ & \ \ $B_3$     & $2.02$      &  $0.080$    \ \ & \ \ $C_3$     &  $2.20$     &  $0.099$    \ \  \tabularnewline
\hline\hline
  \end{tabular}\label{table:CP_GL}
 \end{table}
\end{center}

\begin{center}
 \begin{table}
 \caption{Second critical point location for potentials $A$, $B$ and $C$.}
 \centering{}
 \begin{tabular}{ccc|ccc|ccc}
\hline\hline
\ \ Potential & $T_{c2}^*$  &  $p_{c2}^*$ \ \ & \ \ Potential & $T_{c2}^*$  &  $p_{c2}^*$  \ \ & \ \ Potential & $T_{c2}^*$  &  $p_{c2}^*$ \ \ \tabularnewline \hline
\ \ $A_1$     & $0.34$      &  $4.21$     \ \ & \ \ $B_1$     & $0.35$      & \ $3.44$     \ \ & \ \ $C_1$     &  $0.34$     &  $2.59$     \ \ \tabularnewline
\ \ $A_2$     & $0.48$      &  $3.16$     \ \ & \ \ $B_2$     & $0.48$      & \ $1.86$     \ \ & \ \ $C_2$     &  $0.41$     &  $0.78$     \ \ \tabularnewline
\ \ $A_3$     & $0.59$      &  $1.70$     \ \ & \ \ $B_3$     & $0.57$      & \ $0.49$     \ \ & \ \ $C_3$     &  $0.53$     &  $-0.63$    \ \ \tabularnewline
\hline\hline
  \end{tabular}\label{table:CP_LL}
 \end{table}
\end{center}

\begin{center}
 \begin{table}
  \caption{Values of pressure location in the amorphous region for cases $A$, $B$ and $C$.}
  \begin{tabular}{cc|cc|cc}
  \hline\hline
\ Potentials \ &  Values \ \                            & \ \ Potentials \ &  Values \ \                            & \ \ Potentials \ &  Values     \ \  \tabularnewline\hline
\   $A_1$    \ & $-0.90 \lesssim p^* \lesssim 3.70$ \ \ & \ \   $B_1$    \ & $-0.91 \lesssim p^* \lesssim 3.40$ \ \ & \ \    $C_1$   \ &$-0.85\lesssim p^* \lesssim 2.35$ \tabularnewline
\   $A_2$    \ & $-0.30 \lesssim p^* \lesssim 2.80$ \ \ & \ \   $B_2$    \ & $-0.89 \lesssim p^* \lesssim 1.80$ \ \ & \ \    $C_2$   \ &$-0.76\lesssim p^* \lesssim 0.55$ \tabularnewline
\   $A_3$    \ & $-1.05 \lesssim p^* \lesssim 1.53$ \ \ & \ \   $B_3$    \ & $-1.00 \lesssim p^* \lesssim 0.48$ \ \ & \ \    $C_3$   \ & --  \tabularnewline
\hline\hline
  \end{tabular}\label{table:amorphous}
 \end{table}
\end{center}

At very low temperatures the system becomes less diffusive and
crystallization might be expected. Here we do not explore all the
possible crystal phases, but instead as the temperature is decreased
from the liquid phase an amorphous phase is formed. The dotted line in
Fig.~\ref{fig:PT} shows the
separation between the fluid phase and the amorphous region. The
amorphous phase is identified by the region where the diffusion
coefficient becomes zero and the radial distribution functions do not
exhibit the periodicity of a solid.  Table~\ref{table:amorphous}
 have the characteristic pressure and temperature
values of the amorphous phase boundary for
the different models. It
shows that the region
in the pressure-temperature phase
diagram where the amorphous phase  is present 
shrinks as the shoulder part of the
potential becomes deeper. 

\subsection*{\label{density} Density Anomaly}


Next, we test the effects that the decrease of 
 the distance between the two length scales
and the increase of the depth of the first scale have in the presence of
the density anomaly.  Fig.~\ref{fig:PT} shows the isochores $0.30
\leq \rho^*\leq 0.65$ represented by thin solid lines for
the nine models.  Equation
\begin{equation}
  \left( \frac{\partial V}{\partial T} \right)_{p} = -\left(
  \frac{\partial p}{\partial T} \right)_{V} \left( \frac{\partial
    V}{\partial p} \right)_{T}
\label{tmd}
\end{equation}
indicates that the temperature of minimum
pressure at constant density is the temperature
of maximum density at constant pressure, the TMD. The TMD is the 
boundary of the region of
thermodynamic anomaly, where a decrease in the temperature at constant
pressure implies an anomalous increase in the density and therefore an
anomalous behavior of density (similar to what happens in
water). Figs.~\ref{fig:PT} show the TMD as  solid thick lines. For all
 potentials $A$, $B$ and the potentials $C_1$ and $C_2$ the
TMD is present but for potential $C_3$ the TMD is not observed.

Similarly to what happens with the location of amorphous region, as
the distance between the shoulder and the attractive scales
decreases, the
region in the pressure-temperature phase diagram delimited by the TMD
line shrinks and disappears for the case $C_3$. For the potential $B$ the
TMD line is located at temperatures bellow the temperature of the
liquid-liquid critical point. The thermodynamic parameters that limits
the TMD in phase diagram are shown in the Table~\ref{table:TMD}, where
$p_l$ represents the values of $(\rho^*,T^*,p^*)$ for the point of the
lowest pressure in the TMD line, $p_m$ is the point with the highest
temperature and $p_h$ is the point with the highest pressure.

How can this result be understood?  The density anomalous behavior
arises from the competition between the two length scales: the
shoulder scale and the attractive scale. At high pressures the
shoulder scale wins and at low pressures the attractive scale
wins. The density anomalous region exists only in the intermediate
pressure range where clusters of both scales are present. The value of
the ``high'' pressure and of the ``low'' pressure is determined by the
difference in energy between the two scales. If the difference is too
small the low and high pressures are too close and no anomaly appears.

\begin{table}
 \caption{Limiting values for density ($\rho^*$), temperature ($T^*$) and 
pressure ($p^*$) of the thermodynamics anomalies on pressure-temperature 
diagram. Here the point $p_l$ represents
  the density, temperature and pressure of  the point of the lowest 
pressure in the TMD line, $p_m$ represents the point of the highest 
temperature  and $p_h$ represents the point of the
  highest pressure of the TMD line.}\label{table:TMD}
 \begin{centering}
  \begin{tabular}{ccccc|ccccc|ccccc}
\hline\hline
\ \ cases &          & $p_l$ \ &\ $p_m$ \ &\ $p_h$ \ &\ \ cases &          & $p_l$ \ &\ $p_m$ \ &\ $p_h$ \ &\ \ cases &           & $p_l$ \ &\ $p_m$ \ &\ $p_h$ \ \ \tabularnewline\hline
          & $\rho^*$ & $0.47$\ &\ $0.54$\ &\ $0.61$\ &          & $\rho^*$ & $0.47$\ &\ $0.52$\ &\ $0.57$\ &          &  $\rho^*$ & $0.45$\ &\ $0.51$\ &\ $0.55$\ \ \tabularnewline
\ \ $A_1$ & $T^*$    & $0.71$\ &\ $1.00$\ &\ $0.72$\ &\ \ $B_1$ & $T^*$    & $0.71$\ &\ $0.85$\ &\ $0.69$\ &\ \ $C_1$ &  $T^*$    & $0.56$\ &\ $0.69$\ &\ $0.40$\ \ \tabularnewline
          & $p^*$    & $1.74$\ &\ $3.22$\ &\ $4.50$\ &\ \       & $p^*$    & $1.50$\ &\ $2.50$\ &\ $3.30$\ &\ \       &  $p^*$    & $0.79$\ &\ $1.96$\ &\ $2.60$\ \ \tabularnewline\hline

          & $\rho^*$ & $0.46$\ &\ $0.51$\ &\ $0.59$\ &          & $\rho^*$ & $0.46$\ &\ $0.50$\ &\ $0.54$\ &          &  $\rho^*$ & $0.46$\ &\ $0.48$\ &\ $0.50$\ \ \tabularnewline
\ \ $A_2$ & $T^*$    & $0.70$\ &\ $1.00$\ &\ $0.54$\ &\ \ $B_2$ & $T^*$    & $0.67$\ &\ $0.76$\ &\ $0.63$\ &\ \ $C_2$ &  $T^*$    & $0.49$\ &\ $0.53$\ &\ $0.48$\ \ \tabularnewline
          & $p^*$    & $1.36$\ &\ $2.20$\ &\ $3.17$\ &\ \       & $p^*$    & $0.90$\ &\ $1.40$\ &\ $1.80$\ &          &  $p^*$    & $0.23$\ &\ $0.45$\ &\ $0.64$\ \ \tabularnewline\hline

          & $\rho^*$ & $0.45$\ &\ $0.48$\ &\ $0.54$\ &          & $\rho^*$ & $0.40$\ &\ $0.42$\ &\ $0.43$\ &          & $\rho^*$  &   -     &    -     &   -        \tabularnewline
\ \ $A_3$ & $T^*$    & $0.66$\ &\ $0.89$\ &\ $0.69$\ &\ \ $B_3$ & $T^*$    & $0.44$\ &\ $0.54$\ &\ $0.52$\ &\ \ $C_3$ & $T^*$     &   -     &    -     &   -        \tabularnewline
          & $p^*$    & $0.91$\ &\ $1.27$\ &\ $1.72$\ &          & $p^*$    & $0.15$\ &\ $0.29$\ &\ $0.36$\ &          & $p^*$     &   -     &    -     &   -      \tabularnewline
	  \hline\hline
  \end{tabular}\par
 \end{centering}
\end{table}

\subsection*{\label{diffusion} Diffusion anomaly}


Then, we check  the effects that
the decrease of the distance between the 
two scales have 
 in the location  in the pressure-temperature phase 
diagram of the diffusion
anomaly. The diffusion
coefficient is obtained from the expression:
\begin{equation}
  D = \lim_{t \rightarrow \infty} \frac {\langle \left[ \vec{r}_j (t_0
      + t) - \vec{r}_j(t_0) \right]^2 \rangle_{t_0}} {6t}
  \label{eq:diffusion}
\end{equation}
where $\vec{r}_j(t)$ are the coordinates of particle $j$ at time $t$,
and $\langle \cdots \rangle_{t_0}$ denotes an average over all
particles and over all $t_0$.

\begin{figure}[ht]
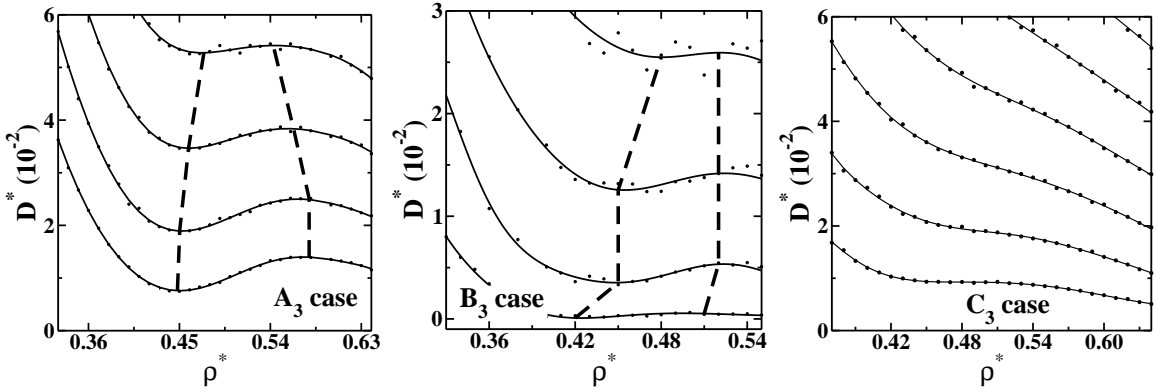

 \begin{centering}
  \begin{tabular}{ccc}
   \includegraphics[clip=true,width=5cm]{dif_A3.eps} &
   \includegraphics[clip=true,width=5cm]{dif_B3.eps} &
   \includegraphics[clip=true,width=5cm]{dif_C3.eps} \tabularnewline
  \end{tabular}\par
 \end{centering}
 \caption{Diffusion coefficient versus reduced  density for
$T^*=0.6,0.7,0.8,0.9$ from the bottom to the
top for the case $A_3$, $T^*=0.4,0.5,0.6,0.7$ from the bottom to the
top for the case $B_3$ and $T^*=0.5,0.6,0.7,0.8,0.9$ from the bottom to the
top for the case $C_3$. The dots are
   the simulational data and the solid lines are polynomial fits. The
   dashed lines connect the densities of minima and maxima diffusivity
   that limit the diffusion anomalous region.}\label{fig:dif}
\end{figure}

Fig.~\ref{fig:dif} shows the behavior of the dimensionless
translational diffusion coefficient, $D^*$, as function of the
dimensionless density, $\rho^*$, at constant temperature for the
cases.  $A_3$, $B_3$ and $C_3$. The solid lines are a polynomial fits
to the data obtained by simulation (the dots in the
Fig.~\ref{fig:dif}). For normal liquids, the diffusion at constant
temperature increases with the decrease of the density. For the cases
$A_1,A_2,B_1,B_2,B_3,C_1,C_2$ (not shown) and for the cases $A_3$ and
$B_3$ the diffusion has a region in the pressure-temperature phase
diagram in which the diffusion increases with density. This is the
diffusion anomalous region illustrated in Fig.~\ref{fig:PT} as a
dashed line.

Similarly to what happens with the location of the TMD, as the two
length scales becomes closer, the region in the pressure-temperature
phase diagram delimited by the extrema of the diffusion goes to lower
pressures, shrinks and disappears for the case $C_3$.

Fig.~\ref{fig:PT} illustrates
that the region in the pressure-temperature
phase diagram  where the dynamic anomaly occurs
englobes the region where the thermodynamic anomaly is present.  This
hierarchy between the anomalies is observed in simulations
~\cite{Ol08a,Er01,Ne01} and in experiments~\cite{An76} for bulk water.

\subsection*{\label{structural} Structural anomaly}


Now, we test the effects that the decrease of the 
distance between the two length scales have in the location in the
pressure-temperature phase diagram of the structural anomalous region.

\begin{figure}[ht]
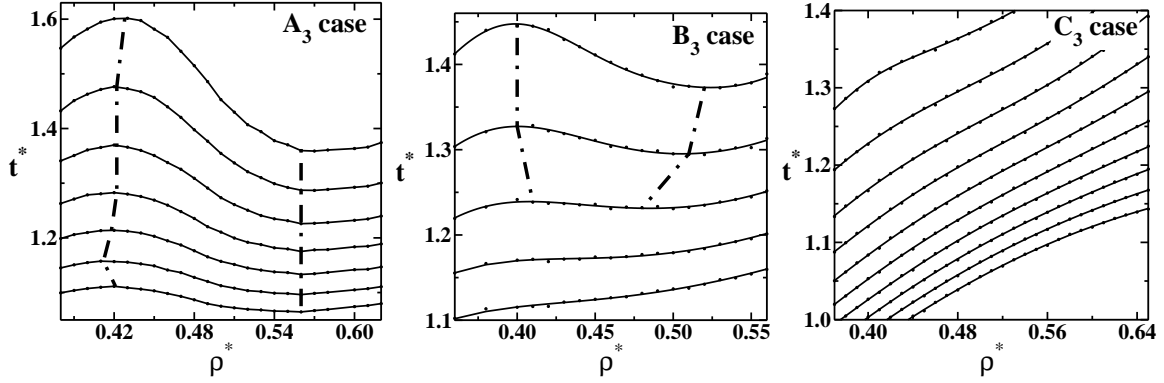

 \begin{centering}
  \begin{tabular}{ccc}
   \includegraphics[clip=true,width=5cm]{tra_A3.eps} &
   \includegraphics[clip=true,width=5cm]{tra_B3.eps} &
   \includegraphics[clip=true,width=5cm]{tra_C3.eps} \tabularnewline
  \end{tabular}\par
 \end{centering}
 \caption{The translational order parameter as a function of 
density for fixed temperatures: 
$T^* = 1.50,\ 1.40,\ 1.30,\ 1.20,\ 1.10,\ 1.00,\ 0.90,\ 0.80,\ 0.70$ 
and $0.60$ (from top to
  bottom) for the setting potentials $A$ and $C$; and 
$T^* = 1.00,\ 0.90,\ 0.80,\ 0.70$ and $0.60$ for the 
setting potential $B$. The dot-dashed lines locate the 
density of maxima e minima
  $t^*$.}\label{fig:trho-1}
\end{figure}

The translational order parameter is defined as~\cite{Sh02,Er01,Er03}
\begin{equation}\label{eq_trans}
  t = \int_0^{\xi_c} \left| g(\xi) - 1 \right| d\xi
\end{equation}
where $\xi = r \rho^{\frac{1}{3}}$ is the distance $r$ in units of the
mean interparticle separation $\rho^{-\frac{1}{3}}$, $\xi_{c}$ is the
cutoff distance set to half of the simulation box times~\cite{Ol06b}
$\rho^{-\frac{1}{3}}$, $g(\xi)$ is the radial distribution function
which is proportional to the probability of finding a particle at a
distance $\xi$ from a referent particle. The translational order
parameter measures how structured is the system.
For an ideal gas it is $g = 1$ and $t = 0$, while for 
the  crystal phase it is $g\neq 1$ over long distances
resulting in a large t. Therefore for normal fluids $t$ increases with the
increase of the density.

Fig.~\ref{fig:trho-1} shows the translational order parameter as a
function of the density for fixed temperatures for potentials $A_3$,
$B_3$ and $C_3$. The dots represent the simulation data and the solid
line the polynomial fit to the data. For the potentials $A_3$ and
$B_3$ there are a region of densities in which the translational
parameter decreases as the density increases. A dotted-dashed line
illustrates the region of local maximum and minimum of $t^*$ limiting
the anomalous region. For the potential $C_3$, $t^*$ increases with
the density. No anomalous behavior is observed. The potentials
$A_1,A_2, B_1,B_2,C_1$ and $C_2$ that do show anomalous behavior are
not shown here for simplicity.

Fig.~\ref{fig:PT} shows the structural anomaly for the families $A$,
$B$ and $C$, as dotted-dashed lines. It is observed that the region of
structural anomaly embraces both dynamic and thermodynamic
anomalies. As the distance between the shoulder
and the attractive scales is decreased, the structural
anomalous region in the pressure-temperature phase 
diagram shrinks.

Another measure of the 
anomalous behavior is the 
orientational order parameter \cite{St83},  $Q_6$. This
parameter  is used to get
information about tetrahedral order of the molecules. 
For two length scales spherical symmetric continuous (continuous force)
potentials, 
$Q_6$ \cite{Ol06b,Al06,Ya05} exhibit a 
region of temperatures in which it decreases
with increasing density. The  maximum of $Q_6$
is located in the same region in the 
pressure temperature phase diagram of the maximum of
the translational order parameter \cite{Ol06b}. A similar
behavior is expected for our potential.

In resume, for all the density, diffusion and structural anomalous 
regions in the pressure-temperature phase diagram, as  the two length 
scales becomes closer  the region of the pressure-temperature phase diagram
occupied by the anomalous region shrinks. The same effect is observed
when the shoulder scale becomes deeper \cite{Ba09}.

\subsection*{\label{rdf} Radial distribution function}


What is the origin of the disappearance of the thermodynamic, dynamic
and structural anomalous behavior with the decrease of the distance
between the scales?  In order to answer to this question
the behavior of the radial distribution function
for the nine different potentials  is 
studied. The radial distribution function is
a measure of  the probability of finding a pair of atoms separated by
$r$. This function is defined as
\begin{equation}
 g(r) = \frac{V}{N^2} \left< \sum_{i,j=1}^{N} \delta 
\left[\vec{r}-( \vec{r}_i(t_0 + t) - \vec{r}_j(t_0) ) \right] \right>_{t_0}
  \label{eq:gr}
\end{equation}
where $\vec{r}_j(t)$ are the coordinates of particle $i$ and $j$ at
time $t$, $V$ is volume of system, $N$ is number of particles and
$\langle \cdots \rangle_{t}$ denotes an average over all particles.

\begin{figure}[ht]
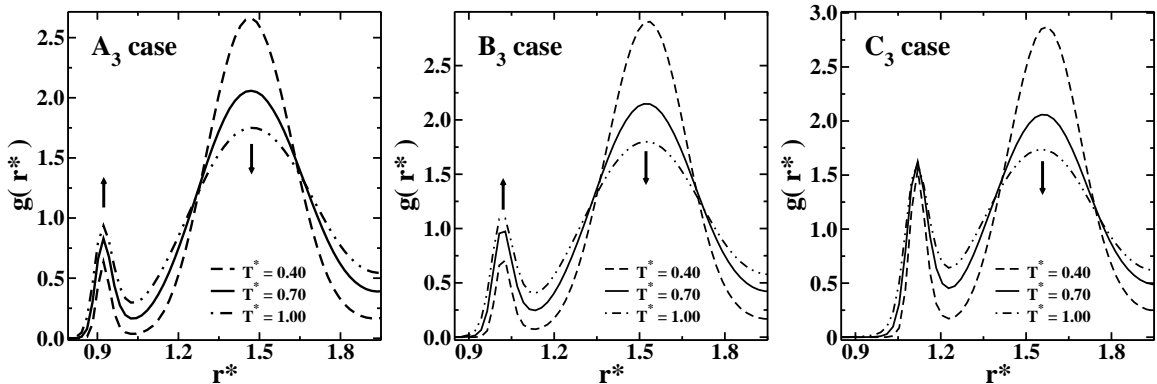

 \begin{centering}
  \begin{tabular}{ccc}
   \includegraphics[clip=true,width=5cm]{gr_0480_A3.eps} &
   \includegraphics[clip=true,width=5cm]{gr_0420_B3.eps} &
   \includegraphics[clip=true,width=5cm]{gr_0390_C3.eps} 
\tabularnewline
  \end{tabular}\par
 \end{centering}
 \caption{Radial distribution function versus reduced distance for
   three cases, $A_3,B_3, C_3$ for reduced density $\rho^*=0.480,
   0.420$ and $0.390$, respectively. In cases $A_3$ and $B_3$ the
   first peak of $g(r^*)$ increases with increasing temperature, while
   the second peak decreases. In $C_3$ case for the first peak
   potential keeps constant while the second peak decreases.}
\label{fig:gr}
\end{figure}

Recently it was shown that a necessary condition for the 
presence of density anomaly is to have
particles moving from one scale to the other as the temperature is
increased \cite{Ol06b,Ol08a,Vi10}, for a fixed density. Here we
test if this assumption is confirmed in the potentials we are
analyzing.  Fig.~\ref{fig:gr} illustrates the radial distribution
function versus distance for a fixed density and various temperatures
for the potentials $A_3$, $B_3$ and $C_3$. For the potentials $A_3$
and $B_3$ the percentage of particles in the first length scale
increases while the percentage of particles in the second length scale
decreases as the temperature is increased. This means that as the
system is heated at constant density particles move from one scale to
the other. This behavior is also observed for the potentials
$A_1,A_2,B_1,B_2, C_1$ and $C_2$ (not shown here for simplicity) and
confirms our assumption that the presence of anomalies is related with
particles moving from one length scale to the other length scale
\cite{Ol06b,Ol08a}. This is not the case for the potential $C_3$. For
the case $C_3$, as the temperature is increased the particles move
from the second to the other further away coordination shells and the
percentage of particles at the first scale is not affected by the
increase in temperature and therefore no anomaly is observed
\cite{Ol06b,Ol08a}.

How can we understand this result? The density anomaly appears if
particles move from the second length scale to the first length scale.
In the case of the potentials $A_3$ and $B_3$ the difference in energy
between the two scales is high and heat is required for having
particles reaching the shoulder length scale.  Consequently, as the
temperature is increased at constant density,  more particles will be at
the first scale and pressure decreases (see Fig.~\ref{fig:gr}). In the
potential $C_3$ the difference in energy between the two length scales
is small. Almost no heat is required to have particles in the first
length scales that saturates. So particles actually do not move from
one scale to the other.

This picture in terms of the presence of particles in the different
shells can also be checked in the framework of the excess entropy
\cite{Ol06b,Ol08a}.

\subsection*{\label{excess} Excess entropy and anomalies}


The excess entropy is defined as the difference between the entropy of
the real fluid and that of an ideal gas at the same temperature and
density.  It also can be given by its two body contribution of
$s_{\mathrm{ex}}$, 
\begin{equation}\label{eq:entropy}
s_{sex} \approx s_2 
=-2\pi\rho\int_{0}^{\infty}{[g(r)\ln g(r) - g(r) +1]r^2dr},
\end{equation}
gives a good approximation of $s_{sex}$.

What can we learn from the excess entropy about the mechanism
responsible for the density, the diffusion and the structural
anomalies?  In the sec.~\ref{rdf} we have shown 
using the radial distribution function that for potentials
that exhibit density, diffusion and structural anomalous behavior as
the temperature is increased particles move from the second
coordination shell to the first coordination shell. 
In the case of
systems in which no anomalies are present, as the temperature is
increased particles will move from the first and second shells to
further shells. Therefore, in principle the information
about the behavior of particles up to the second coordination
shell would be enough to predict if a system would have 
anomalous behavior.

In order to test this assumption we compute the integrals in the
expression for $s_2^{(2)}$ (see Eq.~(\ref{eq:entropy})) up to the
second coordination shell, namely
\begin{equation}\label{eq:entropy2}
s_2^{(2)} 
=-2\pi\rho\int_{0}^{r_2}{[g(r)\ln g(r) - g(r) +1]r^2dr},
\end{equation}
where $r_2$ is the distance of the second shell.

\begin{figure}[ht]
 \begin{centering}
  \begin{tabular}{ccc}
   \includegraphics[clip=true,width=5cm]{s_A3-2.eps} &
   \includegraphics[clip=true,width=5cm]{s_B3-2.eps} &
   \includegraphics[clip=true,width=5cm]{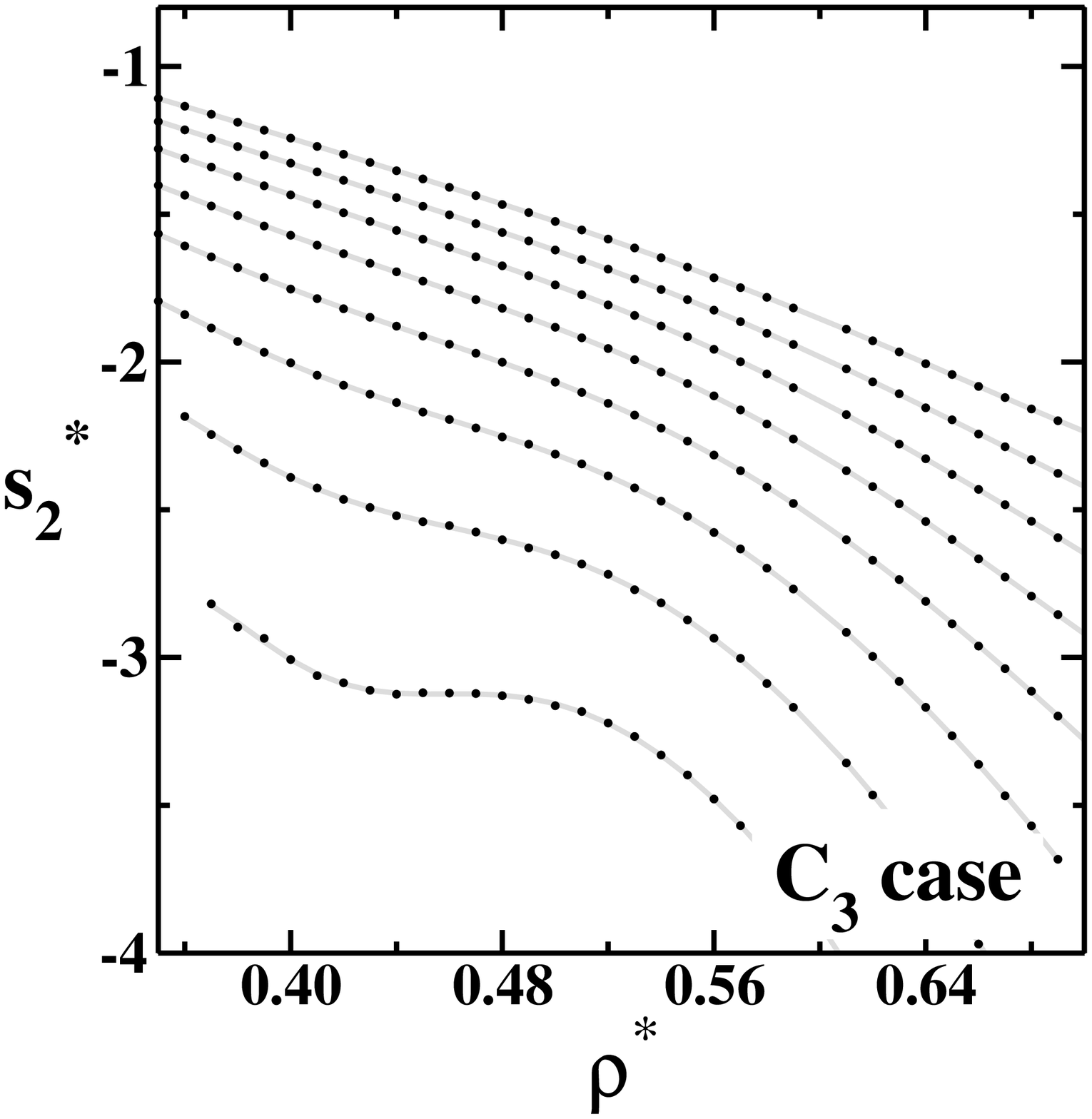} \tabularnewline
  \end{tabular}\par
 \end{centering}
 \caption{Excess entropy, $s_2$,
 versus reduced density for fixed temperatures 
$T^* =0.6,\,0.7,\,0.8,\,0.9,\,1.0,\,1.1$  from the bottom
to the top for the case $A_3$, $T^* =0.6,\,0.7,\,0.8,\,0.9,\,1.0,\,1.1\,,1.2$ 
from the bottom to the top for the case $B_3$ and temperatures 
$T^* =0.5,\, 0.6,\,0.7,\,0.8,\,0.9,\,1.0,\,1.1,\,1.2$  from the bottom
to  the top for the case $C_3$. The temperature $T^*=0.5$ 
contains densities that are metastable points regarding
the high density liquid phase. The
dot-dashed lines locate the density of maxima
 $s_2$ and the dashed, the minima.}\label{fig:entropy}
\end{figure}

Fig.~\ref{fig:entropy} shows the density dependence of $s_2$ along a
series of isotherms spanning from $T^* = 0.60$ to $T^* = 1.50$ for the
cases $A_3$, $B_3$ and $C_3$.

\begin{figure}[ht]
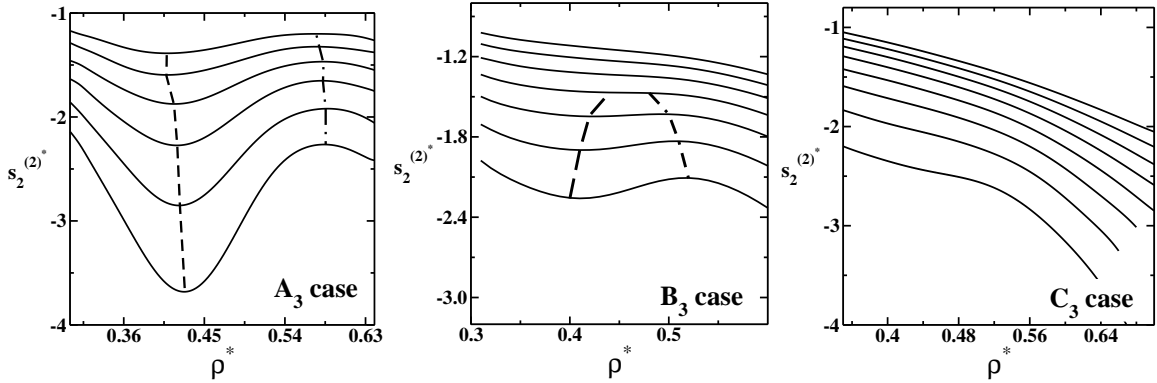

 \begin{centering}
  \begin{tabular}{ccc}
   \includegraphics[clip=true,width=5cm]{s2_A3.eps} &
   \includegraphics[clip=true,width=5cm]{s2_B3-2.eps} &
   \includegraphics[clip=true,width=5cm]{s2_C3-2.eps} \tabularnewline
  \end{tabular}\par
 \end{centering}
 \caption{Excess entropy computed up to the second
coordination shell, $s_2^2$,
 versus reduced density for fixed temperatures 
$T =0.6,\,0.7,\,0.8,\,0.9,\,1.0,\,1.1$  from the bottom
to the top for the case $A_3$, $T =0.6,\,0.7,\,0.8,\,0.9,\,1.0,\,1.1,\,1,2$  from the  bottom
to the top for the $B_3$ and temperatures 
$T =0.5, \,0.6,\,0.7,\,0.8,\,0.9,\,1.0,\,1.1,\,1.2$  from the bottom
to the  top for the case $C_3$. The temperature $T^*=0.5$ 
contains densities that are metastable points regarding
the high density liquid phase. The
dot-dashed lines locate the density of maxima
 $s_2$ and the dashed, the minima.}
\label{fig:entropy2}
\end{figure}

Fig.~\ref{fig:entropy2} shows the density dependence of $s_2^{(2)}$
along a series of isotherms spanning from $T^* = 0.60$ to $T^* = 1.50$
for the cases $A_3$, $B_3$ and $C_3$. The dots are the simulational
data and the solid lines are polynomial fits.

Notice that both $s_2^{(2)}$ and $s_2$ has a maximum and a minimum for
the cases $A_3$ and $B_3$ but not for the case $C_3$ what indicates an
anomalous behavior in the excess entropy.  Comparison between
Fig.~\ref{fig:entropy} and Fig.~\ref{fig:entropy2} shows that the
excess entropy computed up to the second shell not only gives the same
trend but also the same density for the maximum and minimum of the
excess entropy.

Errington \emph{et al.} have shown that the density anomaly is given
by the condition $\Sigma_{\mathrm{ex}}= \left(\partial s_{\mathrm{ex}}
/\partial \ln \rho\right)_T>1$ \cite{Er06}.  They have also suggested
that the diffusion anomaly can be predicted by using the empirical
Rosenfeld's parametrization~\cite{Ro99}.  Based on Rosenfeld's scaling
parameters and approximating the excess entropy 
and its derivative by the two body contribution, namely, 
$s_{\mathrm{ex}}\approx s_2$ and $\Sigma_{\mathrm{ex}}=\Sigma_2$
the anomalous behavior of 
the thermodynamic and dynamic quantities are  observed if
\begin{eqnarray}
excess entropy &\rightarrow &  {\sum}_2 \geq 0  \nonumber \\
diffusivity &\rightarrow &  {\sum}_2 \geq 0.42 \nonumber \\
viscosity &\rightarrow & {\sum}_2 \geq 0.83 \nonumber \\
density &\rightarrow &  {\sum}_2 \geq 1.00 \;\;.
\label{eq:criteria}
\end{eqnarray}

This sequence of anomalies is consistent with the studies of Yan et
al.~\cite{Ya05, Ya06,Ol08a, Ol08b,Si10,Vi10} where structural anomalies are found to
precede diffusivity anomalies, which in turn precede  density
anomalies.

In order to check these criteria in our family of potentials we
compute of $\sum_2$ given by:
\begin{equation}\label{eq:derivative_entropy}
{\sum}_2 = \left(\frac{\partial s_2}{\partial \ln\rho} \right) _T = s_2 - 
2\pi\rho^2\int{\ln g(r) \frac{\partial g(r)}{\partial \rho} r^2dr}\; .
\end{equation}

And also to test if the criteria given by Eq.~(\ref{eq:criteria}) can
also be applied for computations of the excess entropy derivative
computed up to the second shell we also calculate

\begin{equation}\label{eq:derivative_entropy2}
{\sum}_2^{(2)} = \left(\frac{\partial s_2}{\partial \ln\rho} \right) _T = s_2^{(2)} -
2\pi\rho^2\int_{0}^{r_2}{\ln g(r) \frac{\partial g(r)}{\partial \rho} r^2dr}
\end{equation}

\begin{figure}[ht]
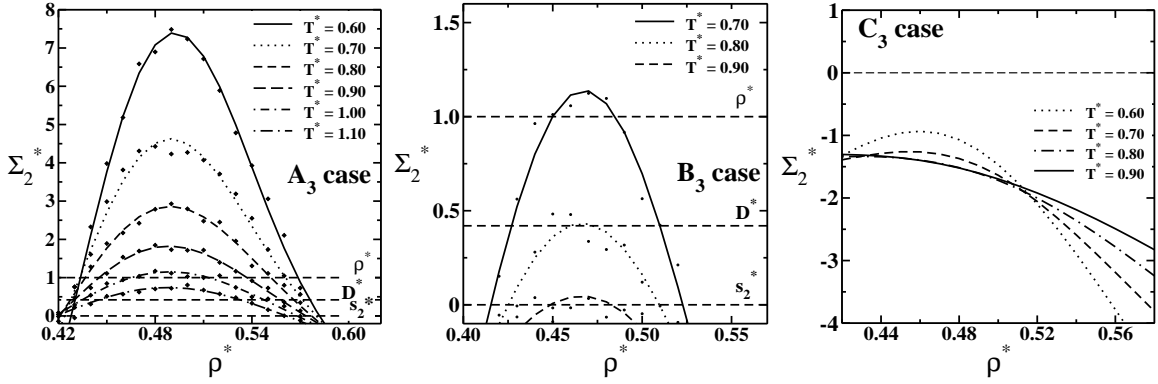

  \begin{centering}
    \begin{tabular}{ccc}
      \includegraphics[clip=true,width=5cm,height=5cm]{ds_A3.eps} &
      \includegraphics[clip=true,width=5cm,height=5cm]{ds_B3.eps} &
      \includegraphics[clip=true,width=5cm,height=5cm]{ds_C3-2.eps}
 \tabularnewline
    \end{tabular}
    \par
 \end{centering}
 \caption{$\sum_2^{(2)}$ versus reduced density
for temperatures  }\label{fig:derivative_entropy}
\end{figure}

\begin{figure}[ht]
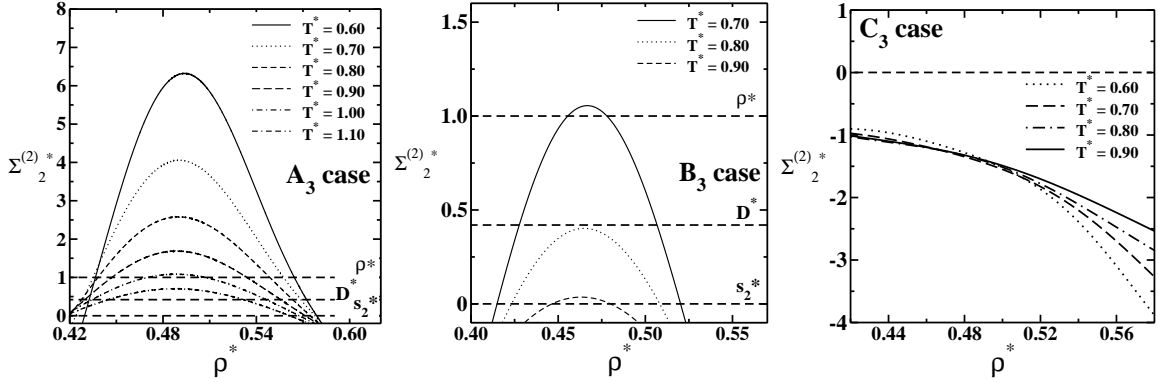

  \begin{centering}
    \begin{tabular}{ccc}
      \includegraphics[clip=true,width=5cm,height=5cm]{ds2_A3.eps} &
      \includegraphics[clip=true,width=5cm,height=5cm]{ds2_B3.eps} &
      \includegraphics[clip=true,width=5cm,height=5cm]{ds2_C3-2.eps} 
\tabularnewline
    \end{tabular}
    \par
 \end{centering}
 \caption{$\sum_2^{(2)}$ versus reduced density  computed up to
   the second coordination shell}\label{fig:derivative_entropy2}
\end{figure}

Fig.~\ref{fig:derivative_entropy} and
Fig.~\ref{fig:derivative_entropy2} show the behavior of $\sum_2$ and
of $\sum_2^{(2)}$ with the density respectively for a fixed
temperature for the potentials $A_3, B_3$ and $C_3$.  The horizontal
lines at $\sum_2 = 0,\ 0.42$ and $1.00$ indicate the threshold beyond
which there are structural, diffusion and density anomalies
respectively.  The graphs confirm that the density, the diffusion and
the structural anomalous behavior is observed for the potentials $A_3$
and $B_3$ but not for the potential $C_3$, confirming Errington's
criteria.

The comparison between Fig.~\ref{fig:derivative_entropy} and
Fig.~\ref{fig:derivative_entropy2} shows that the derivative of the
excess entropy computed up to the second shell is a good approximation
for $\sum_2$ for all the cases.

This result together with the good agreement between $s_2$ and
$s_2^{(2)}$ supports our surmise that focusing in the first and second
shell behavior we can understand the mechanism that leads to the
anomalous behavior.

\section{\label{sec:conclusions} Conclusions}


In this paper we analyzed three families of potentials characterized
by two length scales: a shoulder scale and an attractive scale.  We
found that when approaching the two scales and keeping the slope
between them fixed, the liquid-liquid critical point goes to lower
pressures while keeping the critical temperature fixed. This result
seems to indicate that the slope of the curve might be related to the
critical temperature while the distance between the scales control the
critical pressure. This assumption is also supported 
by another continuous spherical symmetric
potential in which the slope was varied \cite{Si10}.

We also found anomalous behavior in the density, in the diffusion
coefficient, in the structural order parameter and in the excess
entropy for all the cases in which the distance between the scales
were not too short.

From the behavior of the radial distribution function we did infer
that the anomalies are related to particles moving from one scale to
the other.

In order to check our assumption  the excess entropy and
its derivative were computed in two ways: the total
value and the value computed
by integrating up to the second coordination shell. We
found that the behavior obtained by computing these quantities up to
the second coordination shell is accurate both for system with and
without anomalies.   Since the Rosenfeld's
parametrization~\cite{Ro99} is just a lower bound, we can say that it
is enough to compute $s_2$ and $\sum_2$ up to the second shell to know
if the anomalies are present or not.

\section*{ACKNOWLEDGMENTS}

We thank for financial support the Brazilian science agencies CNPq and
Capes. This work is partially supported by CNPq, INCT-FCx.

\end{document}